\documentclass{elsart}

\usepackage{cite}

\usepackage{graphicx}

\begin{document}

\begin{frontmatter}

\title{Accurate calculation of the complex eigenvalues of the Schr\"{o}dinger
equation with an exponential potential}

\author{Paolo Amore \thanksref{PA}}

\address{Facultad de Ciencias, Universidad de Colima, Bernal D\'iaz del
Castillo 340, Colima, Colima, Mexico}

\thanks[PA]{paolo.amore@gmail.com}

\author{Francisco M. Fern\'{a}ndez \thanksref{FMF}}

\address{INIFTA\ (Conicet, UNLP), Divisi\'{o}n Qu\'{i}mica Te\'{o}rica,\\
Diag. 113 y 64 (S/N), Sucursal 4, Casilla de Correo 16,\\
1900 La Plata, Argentina}

\thanks[FMF]{e--mail: fernande@quimica.unlp.edu.ar (corresponding author)}

\begin{abstract}
We show that the Riccati--Pad\'{e} method is suitable for the calculation of
the complex eigenvalues of the Schr\"{o}dinger equation with a repulsive
exponential potential. The accuracy of the results is remarkable for
realistic potential parameters.
\end{abstract}

\end{frontmatter}

\section{Introduction \label{sec:intro}}

Recently, Rakityansky et al \cite{RSE07} developed a method for the
calculation of bound states and resonances of quantum--mechanical problems.
It is based on the approximation of the S--matrix by means of Pad\'{e}
approximants and the location of their poles. The authors mention that the
methods for the calculation of bound states and resonances are usually
developed separately in spite of the fact that those quantities share the
same mathematical nature.

Some time ago we developed the Riccati--Pad\'{e} method that applies to
bound states and resonances of separable quantum--mechanical models \cite
{FMT89a,FMT89b,F92,FG93,F95,F95b,F95c,F96,F96b,F97}. Although the RPM is
less general than the approach proposed by Ratikyansky et al \cite{RSE07} it
is nevertheless an interesting approach for comparison purposes and
benchmark.

The RPM \cite{FMT89a,FMT89b,F92,FG93,F95,F95b,F95c,F96,F96b,F97} is based on
a rational approximation to a modified logarithmic derivative of the
wavefunction that satisfies a Riccati equation. From the coefficients of the
Taylor expansion of this logarithmic derivative, which are functions of the
energy, we construct Hankel determinants. Their roots give rise to sequences
that converge towards the bound states and resonances of the
quantum--mechanical model as the determinant dimension increases. In most
cases the rate of convergence is so great that the RPM yields extremely
accurate real and complex eigenvalues.

The rational approximation and the Hankel quantization condition appear to
select square integrable functions and incoming or outgoing waves. Both,
bound states and resonances emerge from the Hankel sequences because one
does not introduce the boundary conditions at infinity explicitly.

Earlier results for exponential potentials suggest that the RPM may not
yield virtual states and that the Hankel sequences converge to a wrong
limit, although, suspiciously close to the right answer, in the case of some
resonances\cite{F96}. The purpose of this paper is to investigate this
feature of the RPM more closely.

In Section~\ref{sec:model} we discuss a simple model that enables us to
calculate the poles of the scattering amplitude from the roots of Bessel
functions. In Section~\ref{sec:Results} we apply the RPM to this model and
compare approximate and exact results. Finally, in Section~\ref
{sec:conclusions} we summarize our results and draw conclusions.

\section{Model \label{sec:model}}

In this paper we test the performance of the RPM on the Schr\"{o}dinger
equation
\begin{equation}
\psi ^{\prime \prime }(x)+\left( E-Ae^{-\alpha x}\right) \psi (x)=0
\label{eq:Schro1}
\end{equation}
with the boundary condition $\psi (0)=0$. This model has proved useful in
the past for the study of bound, resonance, and virtual states\cite
{M46,ALJ82,MAO93}. Besides, the exponential potential is a suitable
representation of repulsive molecular interactions\cite{ALJ82}. The change
of variables $q=\alpha x$, $\Phi (q)=\psi (q/\alpha )$ leads to an
eigenvalue equation with just one potential parameter:
\begin{equation}
\Phi ^{\prime \prime }(q)+\left( \epsilon -\lambda e^{-q}\right) \Phi (q)=0
\label{eq:Schro2}
\end{equation}
where $\epsilon =E/\alpha ^{2}$ and $\lambda =A/\alpha ^{2}$.

A further change of variables $z=2\sqrt{-\lambda }e^{-q/2}$, $Y(z)=\Phi
(2\ln [2\sqrt{-\lambda }/z])$ transforms the eigenvalue equation (\ref
{eq:Schro2}) into the Bessel equation
\begin{equation}
z^{2}Y^{\prime \prime }(z)+zY^{\prime }(z)+\left( z^{2}-\nu ^{2}\right)
Y(z)=0  \label{eq:Bessel}
\end{equation}
where $\nu ^{2}=-4\epsilon $. The general solution that satisfies the
boundary condition at $q=0$ is \cite{M46}
\begin{equation}
Y(z)=C\left[ J_{-\nu }\left( 2\sqrt{-\lambda }\right) J_{\nu }(z)-J_{\nu
}\left( 2\sqrt{-\lambda }\right) J_{-\nu }(z)\right]  \label{eq:Y(z)}
\end{equation}

If we assume that the RPM will provide the eigenvalues of those solutions
that satisfy $\lim_{q\rightarrow \infty }\Phi (q)=0$, then it follows from
the behaviour of the Bessel function at origin $J_{\nu }(z)\sim \left(
z/2\right) ^{\nu }/\Gamma (\nu +1)$ that they should be roots of
\begin{equation}
J_{\nu }\left( 2\sqrt{-\lambda }\right) =0  \label{eq:eigen}
\end{equation}
with $\mathrm{Re}(\nu )>0$. The roots of this equation are poles of the
scattering amplitude\cite{ALJ82}.

The modified logarithmic derivative of an eigenfunction of
equation (\ref{eq:Schro2})
\begin{equation}
f(q)=\frac{1}{q}-\frac{\Phi ^{\prime }(q)}{\Phi (q)}  \label{eq:f(q)}
\end{equation}
satisfies the Riccati
equation
\begin{equation}
f^{\prime }(q)+\frac{2f(q)}{q}-f(q)^{2}+\lambda e^{-q}-\epsilon =0
\label{eq:Riccati}
\end{equation}
The Taylor series about the origin
\begin{equation}
f(q)=\sum_{j=0}^{\infty }f_{j}q^{j},\;f_{0}=0  \label{eq:f_series}
\end{equation}
converges in a neighbourhood of $q=0$ and the coefficients $f_{j}$ depend on
$\epsilon $.

The main assumption of the RPM is that the roots of the Hankel determinants $%
H_{D}^{d}(\epsilon )=0$, with matrix elements $f_{j+j+d-1}$, $i,j=1,2,\ldots
,D$, are suitable approximations to the eigenvalues of the Schr\"{o}dinger
equation (\ref{eq:Schro2})\cite
{FMT89a,FMT89b,F92,FG93,F95,F95b,F95c,F96,F96b,F97}. Here, $D=2,3,\ldots $
is the determinant dimension and $d=0,1,\ldots $. More precisely, we expect
that there exists a sequence of roots $\epsilon ^{[D,d]}$ of the Hankel
determinants that converges to a given eigenvalue of that Schr\"{o}dinger
equation as $D$ increases.

\section{Results and discussion \label{sec:Results}}

Previous applications of the RPM showed that the rate of convergence of the
Hankel sequences is remarkable for both real and complex eigenvalues\cite
{FMT89a,FMT89b,F92,FG93,F95,F95b,F95c,F96,F96b,F97}. However, in the case of
the exponential potential (\ref{eq:Schro1}) the Hankel sequences were found
to converge to a result slightly different from the one given by equation (%
\ref{eq:eigen})\cite{F96}. For example, Table~\ref{tab:convergence} shows
Hankel sequences for $\lambda =0.5$ and $\lambda =2$ and the corresponding
exact results obtained from the quantization condition (\ref{eq:eigen}).
Both Hankel sequences exhibit great convergence rate but they do not
converge towards the expected result. We also appreciate that the
disagreement between the exact and RPM eigenvalues increases as $\lambda $
decreases. In addition to the great convergence rate, the Hankel
determinants exhibit clustering of roots about the limits of the sequences as $D$
increases, which is an indication of satisfactory convergence and meaningful
result. However, those limits do not completely agree with the exact results
given by (\ref{eq:eigen}).

\begin{figure}[H]
\begin{center}
\includegraphics[width=9cm]{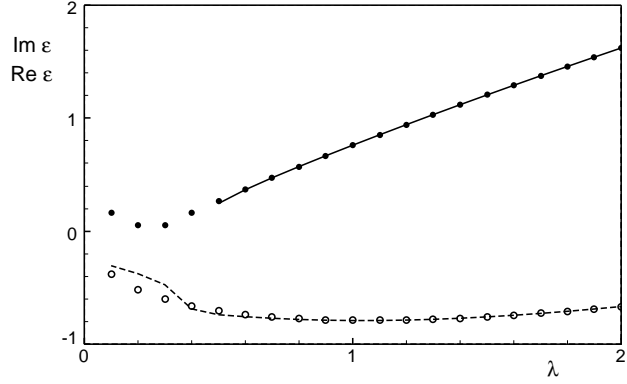}
\end{center}
\caption{Eigenvalues of the exponential potential: $\mathrm{Re}\
\epsilon^{RPM}$ (empty circles), $\mathrm{Im}\ \epsilon^{RPM}$
(solid circles), $\mathrm{Re}\ \epsilon^{exact}$ (dashed line),
$\mathrm{Im}\ \epsilon^{exact}$
(solid line) }
\label{fig:e(lam)}
\end{figure}

\begin{figure}[H]
\begin{center}
\includegraphics[width=9cm]{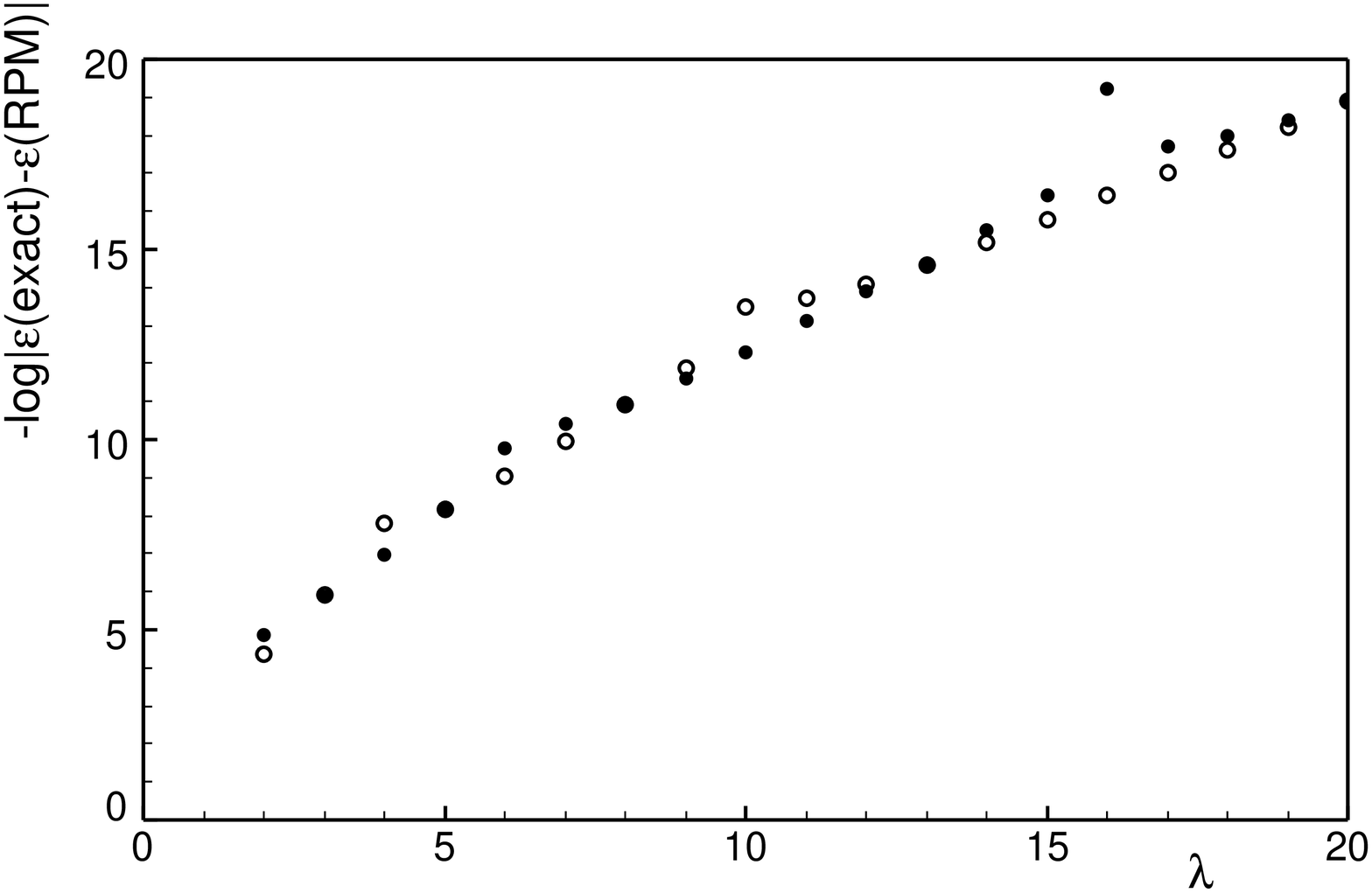}
\end{center}
\caption{$\log|\mathrm{Re}\ \epsilon^{exact}-\mathrm{Re}\ \epsilon^{RPM}|$
(empty circles) and $\log|\mathrm{Im}\ \epsilon^{exact}-\mathrm{Im}\
\epsilon^{RPM}|$ (solid circles) }
\label{fig:loger}
\end{figure}

Fig.~\ref{fig:e(lam)} shows values of $\epsilon (\lambda )$ calculated by
means of the RPM and equation (\ref{eq:eigen}). As indicated above, the
agreement between the RPM and exact results increases as $\lambda $
increases. For $\lambda \leq 0.4$ the RPM yields complex eigenvalues in
spite of the fact that the exact ones are real (see, for example, Atabek et
al\cite{ALJ82} for a more detailed discussion of this behaviour). Moreover,
it is well known that the roots $\nu $ of equation (\ref{eq:eigen}) tend to
negative integers as $\lambda \rightarrow 0$\cite{ALJ82,MAO93} but we
clearly see that the RPM eigenvalues do not exhibit this behaviour. However,
it is most interesting that both real and imaginary parts of the RPM
eigenvalues give a reasonable picture of the behaviour of the exact ones for
large and moderate values of $\lambda $ as shown in Fig.~\ref{fig:e(lam)}.
In order to show the increasing agreement between the RPM and exact
eigenvalue with $\lambda $ more clearly, Fig.~\ref{fig:loger} shows $\log
|\epsilon ^{exact}-\epsilon ^{RPM}|$ as a function of $\lambda $.

The discrepancy between the RPM and exact eigenvalues just mentioned is
interesting from a mathematical point of view, but it is not a serious
drawback for physical purposes. It is well known that realistic potentials
require much larger values of $\lambda $ than those in Fig.~\ref{fig:e(lam)}%
\cite{ALJ82}. If, for example we choose $\lambda =100$ the rate of
convergence of the Hankel sequence is much greater than the one in Table~\ref
{tab:convergence} and the limit agrees with the exact result to at least 20
digits (see Table~\ref{tab:large}). Atabek et al \cite{ALJ82} estimated the
potential parameter for the
$^{2}\Sigma^{+}$ repulsive state of BeH to be $\Lambda \simeq 134$ that
corresponds to $\lambda =4489$. In this case the rate of convergence of the
Hankel sequence is even greater and we obtain the exact result to at least
20 digits with $D=15$ and $d=0$ as shown in Table~\ref{tab:large}. It is
worth mentioning that for such large
values of $\lambda $ we find it easier to obtain the complex energies by
means of the RPM than from the roots of the Bessel function.

\section{Conclusions \label{sec:conclusions}}

We have shown that the Hankel sequences converge to a wrong limit that is
close to the resonances of the repulsive exponential potential. At present
we do not know the reason for this discrepancy that decreases as the
potential strength increases. However the RPM eigenvalue as function of the
potential parameter follows the trend of the exact one. Besides, for
realistic potential functions that fit the interaction between molecular
fragments\cite{ALJ82}, the RPM provides remarkably accurate results and it is
probably more accurate than other methods. Unfortunately we are not aware of
results for such great values of the potential strength, probably because
other methods do not provide the complex eigenvalues so efficiently as the
RPM.

We should mention that the complex eigenvalues of the Schr\"{o}dinger
equation with the exponential potential (\ref{eq:Schro1}) are not what
physicist use to call resonances because the imaginary parts are too large.
However, there has been some interest in their calculation with the purpose
of reconstruction of the scattering amplitude from its poles\cite
{ALJ82,MAO93}.

\begin{table}[H]
\caption{Convergence of the Hankel sequences for two values of $\lambda$}
\label{tab:convergence}
\begin{center}
\begin{tabular}{cll}
\hline
\multicolumn{3}{c}{$\lambda=0.5$} \\ \hline
$D$ & \multicolumn{1}{c}{$\mathrm{Re}\ \epsilon$} & \multicolumn{1}{c}{$%
\mathrm{Im}\ \epsilon $} \\ \hline
10 & -0.70545054582805260895 & 0.26816598479688956569 \\
11 & -0.70545056661473611239 & 0.26816596401669675062 \\
12 & -0.70545056816101389381 & 0.26816596478162896896 \\
13 & -0.70545056810407142688 & 0.26816596486012868610 \\
14 & -0.70545056805626480670 & 0.26816596487152171555 \\
15 & -0.70545056805476309518 & 0.26816596487187249577 \\
16 & -0.70545056805499618233 & 0.26816596487156659341 \\
17 & -0.70545056805503003056 & 0.26816596487158015859 \\
18 & -0.70545056805502881287 & 0.26816596487157957123 \\
19 & -0.70545056805502843245 & 0.26816596487157970586 \\
20 & -0.70545056805502836237 & 0.26816596487157971713 \\ \hline
$Exact$ & -0.73985910415959609800 & 0.24527511363052010569 \\ \hline
\multicolumn{3}{c}{$\lambda=2$} \\ \hline
$D$ & \multicolumn{1}{c}{$\mathrm{Re}\ \epsilon$} & \multicolumn{1}{c}{$%
\mathrm{Im}\ \epsilon $} \\ \hline
10 & -0.66695560251365514674 & 1.6211700286378446183 \\
11 & -0.66695559703717080232 & 1.6211700273918467827 \\
12 & -0.66695559708065060107 & 1.6211700264998970342 \\
13 & -0.66695559710083543597 & 1.6211700264601278937 \\
14 & -0.66695559710850847182 & 1.6211700264500121775 \\
15 & -0.66695559710929174524 & 1.6211700264509149572 \\
16 & -0.66695559710923347745 & 1.6211700264509352898 \\
17 & -0.66695559710922690796 & 1.6211700264509338670 \\
18 & -0.66695559710922683370 & 1.6211700264509344139 \\
19 & -0.66695559710922693659 & 1.6211700264509344652 \\
20 & -0.66695559710922696333 & 1.6211700264509344798 \\ \hline
$Exact$ & -0.66691308506826236505 & 1.6211836543647526877
\end{tabular}
\par
\end{center}
\end{table}

\begin{table}[H]
\caption{Convergence of the Hankel sequences for large values of $\lambda$}
\label{tab:large}
\begin{center}
\begin{tabular}{cll}
\hline
\multicolumn{3}{c}{$\lambda=100$} \\ \hline
$D$ & \multicolumn{1}{c}{$\mathrm{Re}\ \epsilon$} & \multicolumn{1}{c}{$%
\mathrm{Im} \ \epsilon $} \\ \hline
5 & 71.535851840807002875 & 37.763655201763995538 \\
6 & 71.535265703486635320 & 37.763686464878442119 \\
7 & 71.535231112272632786 & 37.763673576025908553 \\
8 & 71.535229874257576952 & 37.763674377953883589 \\
9 & 71.535229860223483406 & 37.763674375421503193 \\
10 & 71.535229855328822688 & 37.763674374272030169 \\
11 & 71.535229855364554354 & 37.763674374318377973 \\
12 & 71.535229855364593976 & 37.763674374316154213 \\
13 & 71.535229855364798246 & 37.763674374316019149 \\
14 & 71.535229855364801073 & 37.763674374316016150 \\
15 & 71.535229855364801145 & 37.763674374316015925 \\
16 & 71.535229855364801148 & 37.763674374316015917 \\
17 & 71.535229855364801148 & 37.763674374316015916 \\ \hline
\multicolumn{3}{c}{$\lambda=4489$} \\ \hline
$D$ & \multicolumn{1}{c}{$\mathrm{Re}\ \epsilon$} & \multicolumn{1}{c}{$%
\mathrm{Im}\ \epsilon $} \\ \hline
5 & 4158.9571348913180566 & 530.18487292989576140 \\
6 & 4158.9534557328433593 & 530.18603523832850612 \\
7 & 4158.9533491082024576 & 530.18608879147133066 \\
8 & 4158.9533468989068860 & 530.18609014361891762 \\
9 & 4158.9533468479121394 & 530.18609015967783227 \\
10 & 4158.9533468461423605 & 530.18609015977128983 \\
11 & 4158.9533468460962321 & 530.18609015979497844 \\
12 & 4158.9533468460956664 & 530.18609015979556450 \\
13 & 4158.9533468460956584 & 530.18609015979556963 \\
14 & 4158.9533468460956581 & 530.18609015979556941 \\
15 & 4158.9533468460956580 & 530.18609015979556943 \\
16 & 4158.9533468460956580 & 530.18609015979556943 \\ \hline
\end{tabular}
\par
\end{center}
\end{table}

\begin{thebibliography}{99}
\bibitem{RSE07}  S. A. Rakityansky, S. A. Sofianos, and N. Elander, J. Phys.
A 40 (2007) 14857-14869.

\bibitem{FMT89b}  F. M. Fern\'{a}ndez, Q. Ma, and R. H. Tipping, Phys. Rev.
A 40 (1989) 6149-6153.

\bibitem{FMT89a}  F. M. Fern\'{a}ndez, Q. Ma, and R. H. Tipping, Phys. Rev.
A 39 (1989) 1605-1609.

\bibitem{F92}  F. M. Fern\'{a}ndez, Phys. Lett. A 166 (1992) 173-176.

\bibitem{FG93}  F. M. Fern\'{a}ndez and R. Guardiola, J. Phys. A 26 (1993)
7169-7180.

\bibitem{F95c}  F. M. Fern\'{a}ndez, Phys. Lett. A 203 (1995) 275-278.

\bibitem{F95}  F. M. Fern\'{a}ndez, J. Phys. A 28 (1995) 4043-4051.

\bibitem{F95b}  F. M. Fern\'{a}ndez, J. Chem. Phys. 103 (1995) 6581-6585.

\bibitem{F96}  F. M. Fern\'{a}ndez, J. Phys. A 29 (1996) 3167-3177.

\bibitem{F96b}  F. M. Fern\'{a}ndez, Phys. Rev. A 54 (1996) 1206-1209.

\bibitem{F97}  F. M. Fern\'{a}ndez, Chem. Phys. Lett 281 (1997) 337-342.

\bibitem{M46}  S. T. Ma, Phys. Rev. 69 (1946) 668.

\bibitem{ALJ82}  O. Atabek, R. Lefebvre, and M. Jacon, J. Phys. B 15 (1982)
2689-2701.

\bibitem{MAO93}  P. Midy, O. Atabek, and G. Oliver, J. Phys. B 26 (1993)
835-853.
\end{thebibliography}
\end{document}